\documentclass[
    ,final            % use final for the camera ready runs
  ]
  {aipproc}

\layoutstyle{6x9}

\usepackage{epsfig,xspace}
\usepackage{color}

\def\NIMA{{Nucl. Instrum. Methods} A}

\def\PRL{Phys. Rev. Lett.}
\def\PRD{{Phys. Rev.} D}

\definecolor{Red}{rgb}{1,0,0}
\definecolor{Blue}{rgb}{0,0,1}
\definecolor{Green}{rgb}{0,1,0}

\def\be{\begin{equation}}
\def\ee{\end{equation}}
\def\bea{\begin{eqnarray}}
\def\eea{\end{eqnarray}}

\def\mes{\ensuremath{m_{\mathrm{ES}}}\xspace}
\def\deltae{\ensuremath{\Delta E}\xspace}
\def\deltat{\ensuremath{\Delta t}\xspace}
\def\dmd{\ensuremath{\Delta m_{\mathrm{d}}\xspace}}
\def\brec{\ensuremath{B_{\mathrm{REC}}\xspace}}
\def\btag{\ensuremath{B_{\mathrm{TAG}}\xspace}}

\def\ifb{\ensuremath{\mathrm{fb}^{-1}}\xspace}

\def\CP{{\rm CP}\xspace}

\def\BR{{\cal B}\xspace}
\def\B{\ensuremath{B}\xspace}
\def\Bz{\ensuremath{B^0}\xspace}
\def\Bzb{\ensuremath{\overline{B}^0}\xspace}

\newcommand\vud {\ensuremath{V_{\mathrm{ud}}}\xspace}

\newcommand\vub {\ensuremath{V_{\mathrm{ub}}}\xspace}

\newcommand\vtd {\ensuremath{V_{\mathrm{td}}}\xspace}

\newcommand\vtb {\ensuremath{V_{\mathrm{tb}}}\xspace}
\newcommand{\e}      [1]   {{\ensuremath{\times 10^{{#1}}}}}
\newcommand{\su}      [1]   {{\ensuremath{SU({#1})}}}

\usepackage{relsize}
\def\babar{\mbox{\slshape B\kern-0.1em{\smaller A}\kern-0.1em
    B\kern-0.1em{\smaller A\kern-0.2em R}}\xspace}

\begin{document}

\title{\boldmath The measurement of $\alpha$ from the B-factories}


\author{A.~J.~BEVAN}{
  address={Department of Physics, Queen Mary University of London, Mile End Road, London E1 4NS, England.}
}

\begin{abstract}
Significant progress toward measuring the CKM angle $\alpha$ has been made by the B-factories over the past decade.
This work has culminated in a constraint on $\alpha$ with a precision of less than $4^\circ$.
\end{abstract}

\maketitle

%%%%%%%%%%%%%%%%%%%%%%%%%%%%%%%%%%%%%%%%%%%%
%% MAINMATTER
%%%%%%%%%%%%%%%%%%%%%%%%%%%%%%%%%%%%%%%%%%%%

%
% Experimental work to cite at the end of the proceedings.
%
%\begin{table}[!ht]
%\begin{tabular}{l|cc}\hline\hline
%Channel                    & \babar     & Belle \\ \hline \hline
%$\B^0\to \pi^+\pi^-$       & B.~Aubert {\it et al.}, [hep-ex] arXiv:0807.4226 & H.~Ishino {\it et al.}, PRL 98, 211801 (2007)\\
%$\B^\pm\to \pi^\pm\pi^0$   & B.~Aubert {\it et al.}, [hep-ex] arXiv:0807.4226 & K.~Abe {\it et al.}, hep-ex/0610065\\
%$\B^0\to \pi^0\pi^0$       & B.~Aubert {\it et al.}, [hep-ex] arXiv:0807.4226 & K.~Abe {\it et al.}, hep-ex/0610065\\
%$\B^0\to \rho^+\rho^-$     & B.~Aubert {\it et al.}, PRD 76, 052007 (2007)    & A.~Somov {\it et al.}, PRD 76, 011104 (2007)\\
%$\B^\pm\to \rho^\pm\rho^0$ & B.~Aubert {\it et al.}, PRL 102, 141802 (2009)   & J.~Zhang {\it et al.}, PRL 91, 221801 (2003) \\
%$\B^0\to \rho^0\rho^0$     & B.~Aubert {\it et al.}, PRD-RC 78, 071104 (2008) & C.~C.~Chiang {\it et al.}, PRD(RC) 78, 111102 (2008)\\
%$\B^0\to \pi^+\pi^-\pi^0$  & B.~Aubert {\it et al.}, PRD 76, 012004 (2007)    & A.~Kusaka {\it et al.}, PRD 77, 072001 (2008) \\
%\hline \hline
%\end{tabular}
%\end{table}

%%%%%%%%%%%%%%%%%%%%%%%%%%%%%%%%%%%%%%%%%%%%%
\section{Introduction}
%%%%%%%%%%%%%%%%%%%%%%%%%%%%%%%%%%%%%%%%%%%%%

The Standard Model of Particle Physics (SM) accounts for CP violation via a single complex parameter in the
Cabibbo-Kobayashi-Maskawa $3\times 3$ quark-mixing matrix~\cite{bevan:CKM}.
 It is possible to write down six triangle relations using the unitarity of the CKM matrix.
One of these, the so-called Unitarity Triangle, has internal angles $\alpha$, $\beta$, and $\gamma$\footnote{The
notation $\phi_1$, $\phi_2$, and $\phi_3$ is used by the Belle Collaboration.}, and these can be studied using decays
of $B$ mesons. In these proceedings I summarize the measurement of $\alpha$, where $\alpha \equiv
\arg\left[-\vtd\vtb^*/\vud\vub^*\right]$, and the $V_{\mathrm{ij}}$ are CKM matrix elements for $i$ to $j$ quark mixing
processes.

There are two B-Factories in existence, one at the SLAC National Laboratory, USA, and the other at KEK, Japan. Both are
described elsewhere~\cite{bevan:bfactories}, and they collide beams of $e^-$ and $e^+$ with a center of mass (CM)
energy predominantly corresponding to the $\Upsilon(\mathrm{4S})$ resonance. \babar and Belle have accumulated 425 and
771\ifb of data at the $\Upsilon(\mathrm{4S})$, respectively.  The beam energies are asymmetric, so that the CM is
boosted with respect to the laboratory frame of reference.  Most of the decay products of the $\Upsilon(\mathrm{4S})$
are $B$-$\overline{B}$ pairs. By studying combinations of the decays of neutral and charged $B$ mesons to $hh$ final
states, where $h$ is a $\pi$ or $\rho$ meson, we can measure $\alpha$.

The first step in this measurement is to isolate signal candidates (\brec).  This is done by using two kinematic
variables that are computed using the known beam energy in order to reduce resolution effects from experimental
reconstruction, and the correlation between these variables. These are: \deltae the difference in reconstructed energy
of the $B$ candidate and half of the total beam energy in the CM frame of reference; and $\mes$ which is an effective
invariant mass of the $B$ candidate computed using the beam energy instead of the reconstructed $B$ candidate energy.
Signal events peak at zero in the \deltae\ distribution, and at the $B$ mass in $\mes = 5.28$
$\mathrm{GeV}/\mathrm{c}^2$.

Having identified the \brec\ the next step in the measurement involves computing the proper-time difference \deltat
between the decay of the \brec\ and the other $B$ in the event (\btag).  In order to compute \deltat we use the Lorentz
boost of the CM in the lab frame and the spatial separation between the reconstructed \brec\ and \btag\ vertices.  It
is important to account for the effect of detector resolution on \deltat. This has several sources including the
ability to reconstruct the \brec\ and \btag\ vertices, and the decay in flight of fully or partially reconstructed
intermediate particles in the \brec\ or \btag. The final ingredient required for a time-dependent CP measurement is to
identify the flavor of the \btag. This is done using a flavor tagging algorithm that is able to assign a $B^0$ or
$\overline{B}^0$ flavor tag to \btag\ at the time it decayed with a probability that depends on the \btag\ final state.
The complement of this is called the mistag probability of the event. As the $B^0-\overline{B}^0$ mixing frequency
\dmd\ is well known, we can determine the flavor of the \brec\ at the instant it decayed up to a dilution factor
related to the mistag probability. For CP eigenstate decays we construct a CP asymmetry as
\begin{equation}
 {\cal A}(\deltat) = \frac{R (\deltat) - \overline{R}(\deltat) } { R (\deltat) + \overline{R}(\deltat) } = \eta_{f}S\sin(\dmd\deltat)-C\cos(\dmd\deltat),\nonumber
\end{equation}
where $R$($\overline{R}$) is the decay-rate for \Bz(\Bzb) tagged events, and $\eta_f$ is the CP eigen-value of the
final state. $S$ or $C$ are non-zero if the decay is CP violating.

In the case of $B\to hh$ decays, the interference between $\Bz-\Bzb$ mixing and tree amplitudes results in $S$ being
related to $\alpha$. However it is not straightforward to interpret the measurement of $S$ directly in terms of
$\alpha$ as there are potential loop (penguin) contributions that complicate the issue.  Nevertheless it is possible to
extract constraints on $\alpha$ using \su{2} isospin or \su{3} flavor based relations~\cite{bevan:theory}.

%%%%%%%%%%%%%%%%%%%%%%%%%%%%%%%%%%%%%%%%%%%%%
\section{Measurements of $\alpha$}
%%%%%%%%%%%%%%%%%%%%%%%%%%%%%%%%%%%%%%%%%%%%%

% pi+pi-
%{$\err{5.5}{0.4}{0.3}$}                           & %   [11]  BABAR arXiv:0807.4226
%{$\err{5.1}{0.2}{0.2}$}                           & %   [63]  BELLE Phys. Rev. Lett. 99, 121601 (2007)
% pi+pi0
%{$\err{5.02}{0.46}{0.29}$}                        & %   [13]  BABAR arXiv:0807.4226
%{$\berr{6.5}{0.4}{0.4}{0.5}$}                     & %   [63]  BELLE Phys. Rev. Lett. 99, 121601 (2007)
% pi0pi0
%\blue{$\err{1.83}{0.21}{0.13}$}% [6]  BABAR  B. Aubert et al., arXiv:0807.4226
%\blue{$\err{1.1}{0.3}{0.1}$}   % BELLE       K. Abe et al., hep-ex/0610065

Before the B-Factories started to take data the most promising way to measure $\alpha$ was thought to be through the
study of $\B \to \pi\pi$ decays as the $\pi^+\pi^-$ state has a large branching fraction, $\BR \sim 5.2\e{-6}$.  The
branching fractions of $\B \to \pi^\pm \pi^0$ and $\pi^0\pi^0$ are required to constrain penguin contributions in this
decay.  The branching fraction for $\pi^\pm\pi^0$ is comparable to that for $\pi^+\pi^-$, however the branching
fraction of $\pi^0\pi^0$ is $1.6\e{-6}$ which turns out to be neither small enough, nor large enough to enable us to
strongly constrain penguin pollution in these decays. We now know that there is a significant penguin contamination in
the $\pi^+\pi^-$ mode. Nevertheless useful constraints on $\alpha$ have been obtained as a result of experimental
efforts over the past decade~\cite{bevan:pipi}.

Historically the \babar\ and Belle measurements of $S$ and $C$ in $\B\to \pi^+\pi^-$ have not been in good agreement,
however in recent years the compatibility between results has improved.  Today the results of the two B-Factories are
in agreement with each other giving a world average measurement of $S=-0.65\pm 0.07$ and $C=-0.38\pm 0.06$.  The
constraint obtained on $\alpha$ is limited by our ability to determine the penguin contribution and the region
$20^\circ \leq \alpha \leq 70^\circ$ has been excluded at more than $3\sigma$ using $B\to \pi\pi$ decays.

The method of measuring $\alpha$ using $B\to \rho\rho$ decays is similar to that used for $B\to \pi\pi$ decays.  The
main differences are that $\rho\rho$ decays have a smaller signal to background ratio, and the final state is a \CP
admixture, where the dominant CP even signal component has to be identified in order to extract time-dependent
information related to $\alpha$.  The dominant CP even signal component is extracted from a maximum-likelihood fit that
accounts for the presence of longitudinal and transverse components.  The fraction of longitudinally polarized (\CP
even) signal is close to unity, which simplifies this process.

The precision of the constraint on $\alpha$ obtained using $\rho\rho$ decays has changed over time due to the
constraints placed on penguin amplitudes. The initial measurements from \babar benefited from a large measured
branching fraction of $\B^\pm\to\rho^\pm\rho^0$~\cite{bevan:earlyrralpha}.  This branching fraction is crucial for the
determination of the penguin contribution.  The result of measuring a large branching fraction for $\rho^\pm\rho^0$ is
a 4-fold degenerate solution in $\alpha$ with flat isospin triangles. In 2007 the measured branching fraction for
$\B^\pm\to\rho^\pm\rho^0$ was reduced, resulting in a larger error on $\alpha$.  The latest $\rho^\pm\rho^0$ branching
fraction measurement has increased the world average, which in turn decreases the penguin pollution, and overall error
on alpha~\cite{bevan:lastrhoprhom,bevan:rhozrhoz}.  The current constraint is $\alpha=(92.4^{+6.0}_{-6.5})^\circ$ using
the Gronau-London \su{2} approach.  This is now in good agreement with the \su{3} approach from Beneke {\em et al.}
which gives $\alpha=(89.8^{+7.0}_{-6.4})^\circ$~\cite{bevan:lastrhoprhom}. The difference in central value of these two
interpretations comes from neglecting a small correction from electroweak loop contributions in the Isospin analysis.
Recent searches for $\Bz\to \rho^0\rho^0$ have resulted in \babar\ seeing evidence for a signal, whereas Belle are
currently unable to confirm the existence of this channel~\cite{bevan:rhozrhoz}.  The \babar analysis included a
time-dependent measurement of $S$ and $C$.  It would be possible to reduce the number of ambiguities in the isospin
analysis of $\rho\rho$ decays with a precision measurement of $S$ and $C$ for $B\to \rho^0\rho^0$.

%%%%%%%%%%%%%%%%%%%%%%%%%%%%%%%%%%%%%%%%%%%%%
%\section{\boldmath Other constraints on $\alpha$}
%%%%%%%%%%%%%%%%%%%%%%%%%%%%%%%%%%%%%%%%%%%%%

One important set of measurements that has been performed is the constraint on $\alpha$ using $\B \to \rho\pi$
($\pi^+\pi^-\pi^0$) decays~\cite{bevan:rhopi}.  While the constraint on $\alpha$ obtained through these measurements is
not as strong as that from $\rho\rho$ decays, it should be noted that a high precision update of this mode will be
instrumental in resolving ambiguities inherent in the measurement of $\alpha$.

In addition to the aforementioned decays, there are a number of other decay modes that may provide useful measurements
of $\alpha$ in the future.  These include $B$ meson decays to final states with axial-vector mesons such as $a_1\pi$,
$a_1\rho$, $a_1a_1$, $b_1\pi$, and $b_1\rho$.  While it should be possible to start measuring $\alpha$ from the some of
these modes with the current $B$ factories, any precision measurements would be the remit of a Super Flavor Factory.

%%%%%%%%%%%%%%%%%%%%%%%%%%%%%%%%%%%%%%%%%%%%%
\section{\boldmath Summary}
%%%%%%%%%%%%%%%%%%%%%%%%%%%%%%%%%%%%%%%%%%%%%

The UTfit~\cite{bevan:utfit} and CKM fitter~\cite{bevan:ckmfitter} groups combine information on $\alpha$ using
different statistical methods.  These groups report average values of $\alpha$ using all measurements as $(92.0 \pm
3.4)^\circ$, and $(90.6^{+3.8}_{-4.2})^\circ$, respectively. Figure~\ref{bevan:fig:alpha} shows the constraint on
$\alpha$ obtained by UTfit on combining direct measurements, where $\alpha = (92 \pm 7)^\circ$.  The corresponding
constraint from CKM fitter is $\alpha = (89^{+4.4}_{-4.2})^\circ$.

\begin{figure}[!h]
  \resizebox{8.0cm}{!}{\includegraphics{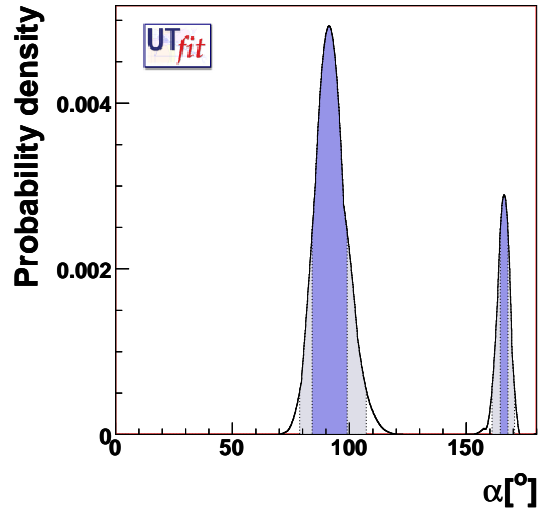}}
 \caption{The UTfit constraint obtained on $\alpha$ using direct measurements.}
\label{bevan:fig:alpha}
\end{figure}

In summary significant progress in the measurement of $\alpha$ has been made over the last decade by the SLAC and KEK
B-Factories.  This provides an important independent crosscheck of the CKM paradigm describing CP violation in the SM.
Future experiments such as the proposed Super Flavor Factories, SuperB in Italy and SuperKEKB in Japan will provide us
with the necessary tools and data to perform precision tests of $\alpha$ in individual channels. Any deviations from SM
expectation measured could hint at new physics corrections to the CKM paradigm.

%%%%%%%%%%%%%%%%%%%%%%%%%%%%%%%%%%%%%%%%%%%%%%%%
%% BACKMATTER
%%%%%%%%%%%%%%%%%%%%%%%%%%%%%%%%%%%%%%%%%%%%%%%%

\begin{theacknowledgments}
The author would like to thank the conference organizers for the opportunity to give this talk, and the scientists
working on the SLAC and KEK B-Factories for their superlative efforts. This work is supported by the UK Science and
Technology Facilities Council.
\end{theacknowledgments}



\bibliographystyle{aipproc}   % if natbib is available

%\bibliography{sample}

\begin{thebibliography}{9}

\bibitem{bevan:CKM}
N.~Cabibbo, \PRL\ {\bf 10}, 531 (1963); M.~Kobayashi and T.~Maskawa, Prog. Theor. Phys. {\bf 49}, 652 (1973).

\bibitem{bevan:bfactories}
\babar\ Collaboration, \NIMA\ {\bf 479}, 1 (2002); Belle Collaboration, \NIMA\ {\bf 479}, 117 (2002).

\bibitem{bevan:theory}
M.~Gronau and D.~London, \PRL\ {\bf 65}, 3381 (1990); M.~Beneke {\it et al.}, Phys.\ Lett.\ B {\bf 638}, 68 (2006).

\bibitem{bevan:pipi}
B. Aubert {\em et al.}, arXiv:0807.4226; K. Abe {\em et al.} \PRL\ {\bf 99}, 121601 (2007).

\bibitem{bevan:earlyrralpha}
B.~Aubert {\it et al.}, \PRL\ {\bf 93}, 231801 (2004); B.~Aubert {\it et al.}, \PRL\ {\bf 91}, 171802 (2003); J.~Zhang
{\it et al.}, \PRL\ {\bf 91}, 221801 (2003).

\bibitem{bevan:latestrhoprhozero}
B.~Aubert {\it et al.}, \PRL\ {\bf 102}, 141802 (2009); A.~Somov {\em et al.}, \PRD\ {\bf 76}, 011104 (2007).

\bibitem{bevan:lastrhoprhom}
B.~Aubert {\it et al.}, \PRD\ {\bf 76}, 052007 (2007).

\bibitem{bevan:rhozrhoz}
B.~Aubert {\it et al.}, \PRD\ {\bf 78}, 071104 (2008); C.~C.~Chaing {\em et al.}, \PRD\ {\bf 78}, 111102 (2008).

\bibitem{bevan:rhopi}
B.~Aubert {\it et al.}, \PRD {\bf 76}, 012004 (2007); A.~Kusaka {\it et al.}, \PRD\ {\bf 77}, 072001 (2008).

\bibitem{bevan:utfit}
The UTfit Collaboration, M.~Bona {\it et al.}, \url{http://www.utfit.org/}.

\bibitem{bevan:ckmfitter}
The CKM fitter Collaboration, J.~Charles {\it et al.}, \url{http://ckmfitter.in2p3.fr/}.

\end{thebibliography}

%  \typeout{******************************************}
%  \typeout{** Please run "bibtex \jobname" to obtain}
%  \typeout{** the bibliography and then re-run LaTeX}
%  \typeout{** twice to fix the references!}
%  \typeout{******************************************}

%%%%%%%%%%%%%%%%%%%%%%%%%%%%%%%%%%%%%%%%%%%
%% The following lines show an example how to produce a bibliography
%% without the help of the BibTeX program. This could be used instead
%% of the above.
%%%%%%%%%%%%%%%%%%%%%%%%%%%%%%%%%%%%%%%%%%%

\end{document}

\endinput
%%
%% End of file `template-6s.tex'.